\pdfoutput=1
\documentclass[11pt]{article}
\usepackage[table]{xcolor}
\usepackage[]{META/eacl2023}
\usepackage{times}
\usepackage{url}
\usepackage{latexsym}
\usepackage[T1]{fontenc}
\usepackage{calc}
\usepackage{tikz}
\usepackage{xkeyval}
\usepackage{ifthen}
\usepackage{hyperref}
\usepackage[normalem]{ulem}
\usepackage{soul}
\usepackage{multirow}
\usepackage{todonotes}
\useunder{\uline}{\ul}{}
\usepackage[utf8]{inputenc}
\usepackage{microtype}
\usepackage{xcolor}
\usepackage{geometry}
\usepackage{times}
\usepackage{latexsym} 
\usepackage{algpseudocode}
\usepackage{algorithm}
\usepackage{amsmath}
\usepackage{mathtools}
\usepackage{cases}
\usepackage[utf8]{inputenc}

\title{HTMOT : Hierarchical Topic Modelling Over Time}
\author{Judicael Poumay \\
  ULiege/HEC Liege\\
  Rue Louvrex 14, 4000 Liege, Belgium \\
  {\tt judicael.poumay@uliege.be} \\\And
  Ashwin Ittoo \\
  ULiege/HEC Liege\\
  Rue louvrex 14, 4000 Liege, Belgium \\
  {\tt ashwin.ittoo@uliege.be} \\}

\date{December 2022}

\begin{document}
\maketitle

\begin{abstract}
Topic models provide an efficient way of extracting insights from text and supporting decision-making.
Recently, novel methods have been proposed to model topic hierarchy or temporality.
Modeling temporality provides more precise topics by separating topics that are characterized by similar words but located over distinct time periods. 
Conversely, modeling hierarchy provides a more detailed view of the content of a corpus by providing topics and sub-topics.
However, no models have been proposed to incorporate both hierarchy and temporality which could be beneficial for applications such as environment scanning.
Therefore, we propose a novel method to perform Hierarchical Topic Modelling Over Time (HTMOT). 
We evaluate the performance of our approach on a corpus of news articles using the Word Intrusion task. 
Results demonstrate that our model produces topics that elegantly combine a hierarchical structure and a temporal aspect.
Furthermore, our proposed Gibbs sampling implementation shows competitive performance compared to previous state-of-the-art methods. 
\end{abstract}
\section{Introduction}
In NLP, over the years, several methods for  extracting themes (or topics) from a corpus have been proposed \citep{Alghamdi2015, Barde2017}. These topic models have been applied to various tasks, including document summarization \citep{YANG20151340}, environment scanning \citep{GREGORIADES2021115546,KIM2020113401}, understanding employee and customer satisfaction \citep{KORFIATIS2019472, BASTANI2019256} among others.

The seminal LDA algorithm \citep{LDA} leads the way for the study of topic models. However, LDA requires the user to specify a predefined number of topics to be extracted. Furthermore, LDA generates a flat topic structure with no hierarchical or temporal information.

Recently, hierarchical topic models have been proposed \citep{nHDP, nCRP}. Such models enable the extraction of topics and sub-topics organised in a tree-like hierarchy. Additionally, these models dynamically determine the appropriate number of topics and sub-topics during training. These models are particularly useful in applications such as ontology learning \citep{8102916} and research idea recommendation\citep{Wang2019}.  

In parallel, temporal topic models have been proposed \citep{ToT,MST,song2008non, DTM}. Incorporating temporal information enables the extraction of topics that can describe events or trends occurring in a corpus. They have been used for tracking trends in scientific articles \citep{10.1145/2020408.2020485} and events in social media \citep{Zhou2013}.  

Intuitively, incorporating temporal and hierarchical information would yield models that encompass the strengths of both. Several applications would benefit from this incorporation such as environment scanning\citep{ELAKROUCHI2021106650}. This task is defined as gathering, analyzing and monitoring information that is relevant to an organization to identify future threats and opportunities. Understandably, this task would benefit from having both hierarchical and temporal modelling. 

Hierarchical modelling would provide more detailed topics as it extracts topics but also sub-topics which deepens our understanding of a thematic. Conversely, temporal modelling would provide more precise topics describing specific events.

However, to date, no topic model integrating both temporal and hierarchical information have been proposed. 
The main reason is the difficulty in integrating time and hierarchy. 
Many temporal topic models have their own structure to represent time, e.g. time trees \citep{MST} or time slices \citep{song2008non, DTM}. Coupling such temporal structure with  a hierarchical structure is extremely challenging. 
Nonetheless, there is one temporal model (ToT) \citep{ToT} that does not require its own structure.  Even in this case, combining time and hierarchy is still difficult for several reasons:
Firstly, the beta distribution used to model time in ToT does not have a known conjugate prior.
Hence, it is not compatible with stochastic variational inference (SVI) used by previous hierarchical models.  
Secondly, applying temporality to every topics would split them into various periods.
Each of these splits would have similar sub-topics, which would lead to an unnecessary multiplication of topics.

Therefore, as our main contribution, we propose a novel method for Hierarchical Topic Modelling Over Time (HTMOT). 
By jointly modelling topic hierarchy and temporality, our model offers the advantages of previous methods, which only focused on a single dimension (i.e. temporality or hierarchy). 
Specifically, we model temporality at the deepest level of the topic tree to extract more precise sub-topics and avoid splitting high level topics.
To the best of our knowledge, our model is the first to jointly model topic hierarchy and temporality. 

As a secondary contribution, we propose a novel implementation of Gibbs sampling. 
We use Gibbs sampling as it was found to be suitable to model temporality \citep{ToT} contrary to SVI. 
However, the original Gibbs sampling implementation is prohibitively slow. 
Thus, we propose an enhanced implementation based on a novel tree-based data structure, which we call the \textit{Infinite Dirichlet Tree}. 
As a result our Gibbs sampling implementation is comparable to SVI in term of speed. 

We performed our experiments using a corpus of 62k news articles and evaluated our method using the Word Intrusion task \citep{readingtea}.

\section{Related Work}
We now describe previous topic modelling methods most closely related to ours. 
Table \ref{tab:related} summarize these models as well as their associated datasets.
For more comprehensive reviews see \cite{Alghamdi2015} and \cite{Barde2017}. 

\subsection{Topic Modelling}
The seminal LDA \citep{LDA} algorithm remains the most popular topic model. 
It is at the basis of most subsequent models. 
At the core of LDA is a Bayesian generative model based on Dirichlet distributions.
These are used to model the document-topic and the topic-word distributions. 
They are learnt and optimized via an inference procedure, which enables topics to be extracted. 
The main weakness of LDA is that it requires the user to specify a predefined number of topics to be extracted. 
However, such information is usually not known in advance. 
Consequently, LDA requires a long model validation step to determine the number of topics. 

The subsequent HDP \cite{HDP} model uses Dirichlet processes (DPs) to determine the number of topics during training.
Using DPs allows us to have an indefinite number of topics contrary to Dirichlet distributions.
Otherwise, HDP operates similarly to LDA. 

\subsection{Hierarchical Topic Modelling}
Methods such as LDA and HDP are only capable of extracting a flat topic structure.
Hence, new methods have been developed to model topic hierarchies.
By extracting topics and sub-topics, we end up with more detailed information about a corpus.

The state-of-the-art for hierarchical topic modelling is nHDP \citep{nHDP}.
It models topic hierarchy by defining a potentially infinite tree where each node corresponds to a topic.
At each branch of the tree, we exactly have the HDP model.
The difference is that, when a word is assigned to a topic during training, there is a chance to go deeper in the tree based on a Bernoulli distribution.
If we do go deeper, we repeat the HDP algorithm with a sub-corpus made up of the documents and tokens assigned to the selected topic.

Other topic models have been proposed to model hierarchy.
hPAM \citep{PAMmix} proposes a directed acyclical graph structure instead of a tree to model topic hierarchy.
Thus, high level topics can share low level topics.
While this provides more precise relationships between topics, it is harder to display and navigate.
LSHTM \citep{LSHTM} recursively applies LDA to the sub-corpus defined by the topics of the previous LDA application.
Hence, each new application of LDA provides a new depth to the topic tree.
However, it requires a pre-defined set of parameters to define the shape of the final topic tree.
Finally, the nCRP \citep{nCRP} is the predecessor of nHDP and works similarly.
Nevertheless, it does not model the document-topic distribution as in nHDP.
Consequently, the extracted documents do not have their own topic tree.
Hence, nHDP is more powerful than LSHTM and nCRP\citep{LSHTM,nCRP} while keeping a strict tree structure contrary to hPAM \citep{PAMmix}.

\subsection{Temporal Topic Modelling}
Previous works also investigated the temporality of topics.
Providing information about when a topic occurred and/or how it evolved.
Understanding the temporality of topics is important, especially for environment scanning where events and changes in the environment are important signals.

The ToT \citep{ToT} model is a modified version of LDA which incorporates temporality.
Each document/word is associated with a timestamp which are used to fit a beta distribution for each topic. 
This beta distribution is optimized jointly as the topics are being discovered.
The results show topics that are either better localized in time (events with specific dates) or with a clear evolution through time (growth/decline).

Other topic models have been proposed to model temporality.
MTT \citep{MST} creates a tree for each topic which provides the ability to understand topics at various time scale.
Specifically, deeper nodes correspond to a smaller timescale.
DTM \citep{DTM} slices the corpus by periods.
The first slice is processed similarly to LDA and the following slices are processed using the previous one as prior.
Finally, DCTM \citep{song2008non} also slices the corpus in period.
However, it uses Gaussian processes and SVD instead of LDA based techniques.
The advantage of ToT is that it is non-Markovian and it models time as a continuum.
Hence, ToT is the only model which does not require its own structure to model time such as slices or a binary tree.
This is important if we are already building a structure for the topic hierarchy.


\subsection{Topic Models Evaluation}
\label{topiceval}
Various methods have been used in previous studies to evaluate topic models such as perplexity and coherence. 
However, these methods have been repeatedly demonstrated to be uncorrelated with human judgement \citep{readingtea, incoherence}. 

The Word Intrusion task is the latest evaluation method devised. 
For each topic, it involves inserting an intruder word in the topic top word list and then asking annotators to find it \citep{readingtea}. 
This intruder is selected at random from a pool of words with low probability in the current topic but high probability in some other topic to avoid rare words. 
The idea is that in good topics, the annotators would easily find this intruder. 
With this evaluation method, the final score corresponds to the average classification accuracy made by humans. 
In \cite{machinereadingtea} , they have shown that this task can be automated with performance similar to human annotators.

\section{HTMOT : Hierarchical Topic Modelling Over Time}
We now describe our method for Hierarchical Topic Modelling Over Time (HTMOT). 
We begin by presenting a new type of data structure at the core of HTMOT (section \ref{IDT}). 
Next, we describe how temporality was incorporated into the hierarchy (section \ref{timemod}).
Then, we detail our novel implementation of Gibbs sampling (section \ref{gibbstraining}). 
Finally, we denote important differences between HTMOT and its predecessor  (section \ref{HTMOTvNHDP}). 
\subsection{Counting words using Infinite Dirichlet Trees}
\label{IDT}
Infinite Dirichlet Trees (IDTs) are efficient tree-based data structures we developed. 
The name refers to the potentially infinite number of topics provided by the Dirichlet Processes, which define how they grow. 
The role of these trees is to model the topics, their hierarchical dependency and temporality.
Hence, these trees are optimized during the training process to serve as the final output of HTMOT.

\begin{figure}[ht]
    \centering
    \includegraphics[width=0.7\linewidth]{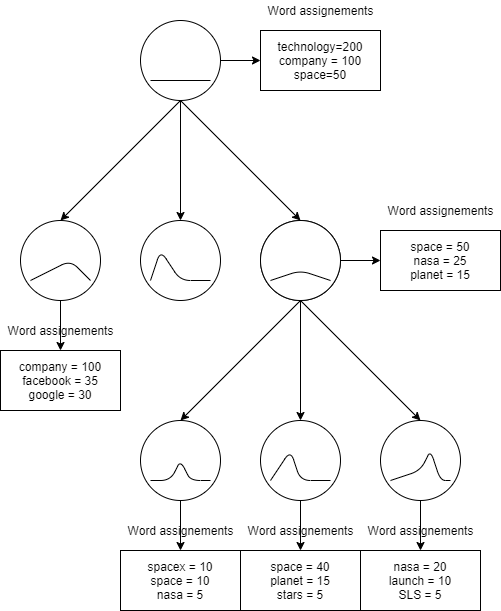}
    \caption{Example of an IDT with word assignments and time distribution (inside nodes). }
    \label{treeex}
\end{figure}

Each node of an IDT is identified by a finite path in the tree as a sequence of node ids, starting from the root.
For example the node "root.A.B" corresponds to a sub-topic of the topic "Root.A".
The nodes record word assignments (see figure \ref{treeex}) and the timestamps of those words (associated with the source document). 
Thus, each node represents a topic and defines a \textit{topic-word} and a \textit{topic-time distribution}.

The trees also model the hierarchical distribution of topics. 
Words are assigned to a final topic and to all ancestors of that  topic. 
Hence, there are two types of word assignments : "through" and "final", respectively for the ancestor topics and final topic.
This creates a hierarchical dependency between the nodes and thus a \textit{hierarchical distribution}. 

We use multiple IDTs, one for the corpus and one for each document. 
All words in the corpus are assigned to nodes of the corpus tree. 
Similarly, each document has an associated document tree recording each word of that document. 
Hence, combining all document trees together would yield the corpus tree. 
For both the corpus and document trees, each node (topic) will be assigned a different number of words.
Thus, nodes differ in size which creates a distribution. 
Hence, the corpus tree defines a \textit{corpus-topic distribution} and each document tree defines a\textit{ document-topic distribution}.

From the foregoing discussion, we can see that the assignment of words to the different trees defines the \textit{topic-word, topic-time, document-topic, corpus-topic and topic-hierarchy distributions.}
Hence, by simply moving words around in those trees, we can optimize all these distributions jointly. 
Once optimized, the trees can be used directly as output to view topics, their hierarchy and temporality for the corpus and each document. 

\subsection{Modelling temporality}
\label{timemod}
Temporality is modelled by associating topics with a beta distribution as in ToT \citep{ToT}. 
This allows us to extract topics that describe specific event in time.
Mathematically, we separate topics that are lexically similar but located at different periods in time.
However, applying temporality to high level topic would split them into various periods.
Each of these splits would have similar sub-topics, which would lead to an unnecessary multiplication of topics.
Hence, contrary to ToT, we do not apply temporality to all topics but only deep ones.
For our experiments, we choose depths of 3 or more. 
This allows us to extract precise topics about specific events in time at the deeper levels while keeping the high level topics intact. 

The parameters of the beta distribution $\rho^1_i$ and $\rho^2_i$ are computed for a topic $i$ based on the current timestamps assignments (associated with each word assignment). 
We used the method of the moment to estimate these parameters :

\begin{equation}\label{timeparam}\rho^1_i = \overline{t_i}*(\frac{\overline{t_i}*(1-\overline{t_i})}{\sigma_{t_i}}-1)\end{equation} 
\begin{equation}\rho^2_i = (1-\overline{t_i})*(\frac{\overline{t_i}*(1-\overline{t_i})}{\sigma_{t_i}}-1)\end{equation}
Where $\overline{t_i}$ is the empirical average timestamp assigned to topic $i$ and $\sigma_{t_i}$ is the empirical variance. 
These parameters are updated each time a word is assigned or unassigned to topic $i$. 

\subsection{Training HTMOT using Gibbs sampling}
\label{gibbstraining}

\begin{algorithm}
		\caption{Traditional Gibbs sampling}
		\label{array-sum}
		\begin{algorithmic}[1]
			\Procedure{ClassicGibbs}{$corpus$}
			\For {N iterations}
    			\For {each $document$ in $corpus$}
    				\For {each $word$ in $document$}
    				\State Sample word-topic assignment
    				\State Sample topic-word
    				\State Sample document-topic
    				\State Estimate time-topic
    				\State Sample corpus-topic
    				\State Sample hierarchy-topic
    			\EndFor
    			\EndFor
			\EndFor
            \State Return solution
			\EndProcedure
		\end{algorithmic}
  \label{tradgibbs}
\end{algorithm}

Two methods are commonly used for training topic models : Gibbs sampling and Stochastic Variational inference (SVI).
Gibbs sampling is asymptotically exact, i.e. it  can exactly approximate the target distribution, unlike SVI \citep{SVIGS}.
However, classical implementations of Gibbs sampling are prohibitively slow as  they require sampling from all distributions (see algorithm \ref{tradgibbs}). 

Nevertheless, in the context of topic modelling, we can avoid this issue \citep{pmlr-v13-xiao10a} and greatly speed up the process. 
Specifically, it is possible to only draw from the word-topic assignment distribution.
This requires the construction of a data structure tailored to the model to implicitly represent the other distributions. 
This is the role played by our Infinite Dirichlet Trees. 

As stated in section \ref{IDT}, IDTs model the aforementioned distributions based on how words are assigned to them. 
Hence, simply by iteratively re-arranging the words in the trees, we are implicitly optimizing these distributions. 
This is the key to speed up the Gibbs sampling process and represents our secondary contribution.

\begin{figure}[ht]
    \centering
    \includegraphics[width=.52\textwidth]{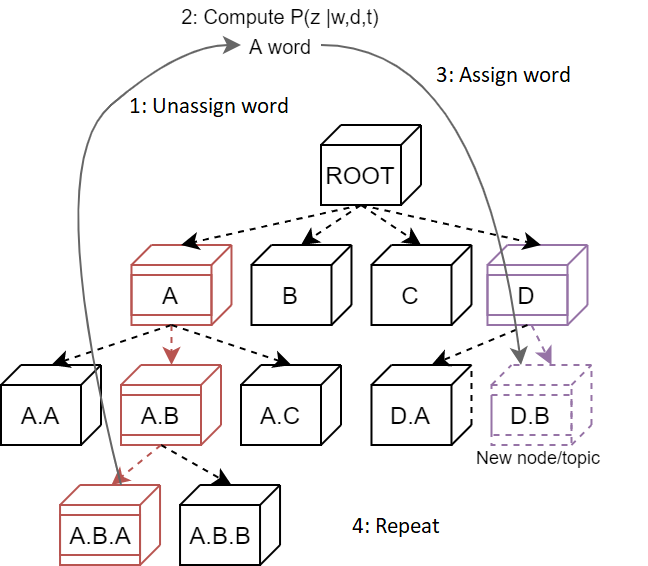}
    \caption{Gibbs sampling with Infinite Dirichlet Trees. Repeat for each word of each document until convergence.}
  \label{gibbspass}
\end{figure}

Hence, our training procedure consists essentially of three steps (see figure \ref{gibbspass}).
For each word of each document in the corpus :
\begin{enumerate}
\item Unassign the word from its current topic (and its ancestors) in the corpus and associated document tree. 
\item Draw a topic assignment for that word from the word-topic assignment distribution. 
\item Re-assign the word to the chosen topic (and its ancestors) in the corpus tree and associated document tree. 
\end{enumerate}
This procedure is repeated  until convergence. 
Note that, changing a word's topic assignment will also update the estimated time parameters of the affected topics (equation \ref{timeparam}).
The initialization procedure of our algorithm is similar expect that it ignores the first step as all words starts unassigned.

\subsubsection{Sampling topic-word assignments (paths in the trees)}
\label{wordass}
We will now explain the procedure behind sampling from the word-topic assignment distribution.
When drawing a topic assignment for a word we have three possible outcomes:
\begin{enumerate}
\item We draw a node/topic from the associated document tree.
\item We draw a node/topic from the corpus tree.
\item We create a new node/topic.
\end{enumerate}

Formally, given a word $w$ with timestamp $t$ in document $d$, we wish to draw a new topic assignment $z$.
As stated  in section \ref{IDT}, topics are identified as a sequence of node ids.
Thus, we iteratively draw the random sequence $z_{0,L} = (z_0,...,z_L)$.
The length $L$ of this sequence is decided by sampling a Bernoulli distribution in-between the sampling of each $z_j$.

Hence each $z_j$ is sampled as :

\[z_j|w,d,t \sim\]
\begin{numcases} {} 
\label{eq:sampling}
with\ probability\ $$\frac{n_d}{\alpha+n_d}$$: \\ 
$$\sum_k \frac{\beta_{k} (t)*(A(k|d)+\epsilon)*(A(k | w)+\phi)*\delta_k}{(A(k)+(\phi*V)) *n_d}$$\\
with\ probability\ $$\frac{n_w}{\beta+n_w}*(\frac{\alpha}{\alpha+n_d})$$ :\\
$$\sum_k \frac{\beta_{k} (t)*(A(k |w)+\phi)*\delta_k}{n_w}$$\\
with\ probability\ $$\frac{\beta}{\beta+n_w}*\frac{\alpha}{\alpha+n_d}$$:\\
new\   topic \label{eq:samplingN}
\end{numcases}

All the variables are explained in table \ref{tab:samplingz} (Appendix).

Note that sampling a node from the corpus tree can lead to the creation of a new node in the associated document tree if that node does not already exist. 
However, when creating an entirely new node, it is created in both trees (corpus tree and associated document tree).

Once a topic $z_j$ is drawn, we draw from a Bernoulli with parameter $p$ to decide if we stop or go deeper in the tree:

\begin{equation}\label{bern1}p=\frac{P + \theta_1}{N + \theta_1+ \theta_2 + C + P}\end{equation}. 
\begin{equation}P = \frac{\beta_{j}(t) *(A^*(z_{0,j}|w) + \phi)*(A^*(z_{0,j}|d) + \epsilon)}{A^*(z_{0,j})+(\phi*V)}\end{equation}
\begin{equation}N = \frac{\phi*\epsilon}{\phi*V}\end{equation}
\begin{equation}\label{bernn}C = \sum_k \frac{\beta_{k}(t)*(A(k|w) + \phi)*(A(k|d) + \epsilon)}{A(k)+(\phi*V)}\end{equation}

All the variables are explained in table \ref{tab:bernoulli} (Appendix).

To summarize, when drawing a topic assignment for a word, we either draw from the document tree, corpus tree, or we create a new topic. 
Then, we draw from a Bernoulli to decide if we go deeper or not. 
If we do go deeper, we repeat the same process until we eventually stop. 
This process is then applied repeatedly too all of the words in the corpus multiple times until convergence.

\subsection{Comparing HTMOT vs. nHDP}
\label{HTMOTvNHDP}
The main difference between HTMOT and nHDP lies in their respective use of the aforementioned Gibbs sampling and SVI training procedures. However, other notable differences exist:
First, our HTMOT algorithm starts with all words unassigned whereas nHDP starts with a pre-clustering step using k-means. 
Second, we do not make use of a greedy algorithm to select trees for each document, i.e, the tree for each document is created automatically as the Gibbs sampler progresses. 
Hence, our training algorithm is thus simpler and easier to implement by avoiding the need for pre-clustering or greedy procedures.

\section{Experimental setup}
\subsection{Dataset}
To perform our experiments, we crawled \footnote{The crawling was performed using Python with the help of the BeautifulSoup library.} 
62k articles from the Digital Trends \footnote{\url{https://www.digitaltrends.com/}.} archives from 2015 to 2020. 
This news website is mainly focused on technological news but also contain general news.
For all articles, we extracted the text, title and timestamp.

The timestamps are mapped to a number between 0 and 1 which corresponds to the domain of the beta distribution used. 
Hence, 0 corresponds to the earliest date of a document in the corpus and 1 corresponds to the latest.

We cleaned the data as follows.
First, we removed common editor's sentences such as "\textit{we strive to help our readers ....}". 
Then, we relied on Spacy's NER and POS to filter relevant tokens. 
Precisely, we kept specific kinds of entities (Person, Norp, Fac, Org, Gpe, Loc, Product, Event, Work\_Of\_Art, Law, Language) and POS elements (ADJ, NOUN, VERB,INTJ, ADV).  
Finally, lemmatization was also applied.

A good pre-processing procedure is essential for the interpretability of topics as shown in \citep{martin-johnson-2015-efficient}.  
Hence, our extraction of named entities aims at enhancing the topics' interpretability by showing actors in the topic such as personalities and companies. 
The training algorithm will not discriminate between words and entities but the visualization interface does.
This means that a topic is no longer displayed as a simple list of words but is instead represented by a list of words and a list of entities.

\subsection{Parameters}
\label{param}
Many parameters control the behavior of our model; this section will describe each of them.

First, we have the Infinite Dirichlet Trees parameters. 
$\alpha$ : the rate at which we create new topics in the document trees. 
$\beta$ : the rate at which we create new topics in the corpus tree. 
$\theta$ : how likely we are to create deeper sub topics. 

Second, we have parameters that regulate the growth of the trees. 
These help speed up the algorithm and keep memory usage to a minimum.
CM (Critical Mass) : the minimum valid size of a topic; only valid topics are part of the final output. 
SM (Splitting Mass) : the minimum size of a topic before it can create sub-topics.
Both are defined as a percentage of the total number of words in the corpus. 
TTL (Time To Live) : how many pass through the corpus before destroying a non-valid node. 
Nodes are also destroyed when they become empty.

Third, we have the Dirichlet prior parameters as in the traditional LDA model. 
$\phi$ : the prior for the topic-word distribution. 
$\epsilon$ : the prior for the corpus and document-topic distributions. 

Finally, we have training parameters. 
Iterations : how many batches we will go through during training. 
SGI (Stop Growth Iteration) : a point at which node new nodes won't be created. 
Set SGI $<$ Iterations to ensure that the last topic to be created has time to converge.

Table \ref{tab:param} defines the value of each parameter used to perform our experiments.

\section{Results and Discussion}
We now present our results, starting with a statistical analysis of the training behavior of HTMOT. 
Then, we will discuss the results of the Word Intrusion task, its drawbacks and directions for future topic modelling evaluation methods. 
Finally, we will examine the various extracted topics qualitatively. 

\subsection{Convergence rate, training speed and algorithmic complexity}

We assessed the convergence of our method by looking at the frequency of topics over time during training.
This frequency indicates how many words in  the corpus are assigned to each topic.
As these frequencies stabilise (the curves flatten), this indicates that the model converged.
However, as hierarchical topic models extract hundreds of topics, observing the frequency of each topic is not reasonable.
Thus, we only observe the convergence of depth 1 topics.

We observed that the convergence rate of our training algorithm is sub-linear with respect to the dataset size. 
These experiments were performed by using samples of the full dataset.
Specifically, using a dataset which is ten time smaller leads to a halving of the time to convergence.
However new topics created during training will perturb this convergence. 
Hence, we prevent this issue with the SGI parameter (see section \ref{param}) which provides a period at the end of training where no topics can be created. 

Now let's consider actual training time of our training procedure with respect to the nHDP's SVI procedure.
Unlike our method (HTMOT), nHDP lacks a temporal component.
Therefore, to ensure a fair comparison in our experiments, we disabled HTMOT's temporal modelling.
We observed that our sampler analyses 135k documents per hour \footnote{Using Python 3.6 with a Ryzen 5 3600x, 32Go RAM and a NVMe SSD.}. 
For nHDP, based  on the figures reported in \cite{nHDP}, we can estimate that SVI analyses roughly 90k articles per hour  \citep{nHDP} \footnote{Using Matlab. 
However no information about hardware was provided}.  
Hence, we believe that our training algorithm is comparable to nHDP's SVI in term of speed.
This observations contradicts previous wisdom that SVI is considerably faster than gibbs sampling \cite{nHDP}.
Overall our model achieved convergence after 10h of training on the full dataset.

We observed that the algorithmic complexity is linear with respect to the dataset size. 
However, the depth of the topic trees and, thus, the growth and regulating parameters for the IDTs can greatly impact performance. 

\subsection{Results of the Word Intrusion task}
\begin{figure*}[ht]
    \centering
    \includegraphics[width=1\textwidth]{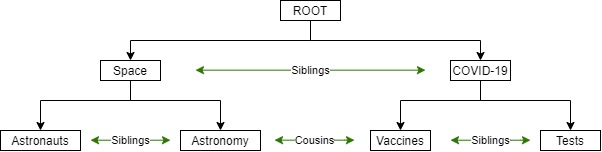}
    \caption{Example of a topic tree with cousins and siblings.}
    \label{cousin}
\end{figure*}

We applied the Word Intrusion task to evaluate our model. 
The original Word Intrusion task involves selecting an intruder word from any other topic. 
However, since our model is hierarchical, we decided to select intruder words from sibling topics only. 
This makes the task more difficult as deeper topics tend to be more lexically related to their siblings. 
This is important as we want topics to be distinct from their siblings. 
Let's take the example of selecting an intruder word for the sub-topic of "astronomy".
In the classical Word Intrusion task, we could choose any topic such as the "Covid-19 vaccines" topic (see figure \ref{cousin}).
In our case, we would choose from one of its siblings such as the "astronaut" topic instead.
Hence, the chosen intruder is semantically much closer to the target topic "astronomy".
Thus, this provides a more robust evaluation of topic quality.

We performed this task using a survey created with Google Forms \footnote{This form is available on github (anonymized link).}. 
The survey required annotators to select an intruder word for each topic presented and provide a confidence score for each answer.
The annotators come from an internet community involved in sharing and answering surveys \footnote{\url{https://www.reddit.com/r/SampleSize/}.}. 
57 respondents answered the survey over the month of may 2021.

Results show 74.83\% accuracy in the Word Intrusion task (as defined in section \ref{topiceval}) on 6 topics at various depths. 
We have also used the automated the word intrusion by replicating the method of  \cite{machinereadingtea}.
We observed an accuracy of 79\%.
Hence, both automated and non-automated methods show results on par with LDA's performance shown in the original Word Intrusion task paper \citep{readingtea}. 

\subsection{Qualitative examination of the resulting topics }
Now, we will inspect a selection of topics to illustrate the capabilities of our HTMOT model. 
Specifically, we will focus on the high level topic of space exploration and its sub-topics. 

Figure \ref{space} presents a depth 1 topic and two of its sub-topics. 
We can clearly see that the parent topic is about space, and the two sub-topics are about astronomy and astronauts , which are indeed related to the parent topic. 
This example also illustrates how entities can help interpret and understand these topics. 
For example, in the astronauts topic, we can see that Bob Behnken, (Doug) Hurley and SpaceX are important entities. 
A quick look at the top documents for that topic show that they were the first to fly on a SpaceX rocket. 
Moreover, in the astronomy topic, Hubble and Spitzer are frequent entities. 
This is coherent as they are two important low earth orbit telescopes. 
Other sub-topics of space include satellite launches, rovers, exoplanets, test flights, etc.

Figure \ref{tempo} shows an example of temporality and hierarchy working together. 
In this representation, we can see the estimated time distributions  (we show the years 2020 and 2021 to have smaller charts).
Here, we have the sub-topic of astronauts and its own three sub-topics: the historic test launch of the spaceX Dragon capsule, the crew 1 launch and the crew 3 launch. These topics were interpreted mostly from top documents. This is because at these depth topics become so small they are difficult to interpret based on top words as they are so precise. 
These topics are depth 3 topics which means temporal modelling is enabled. We can see that they are well localized in time as their associated time distribution is narrow.
The estimated time distribution of the sub-topics matches the timing of the aforementioned events : May 2020, November 2020 and November 2021.
This demonstrates that the ability of our model to extract atomic events at the deeper level of the tree.
Noticeably, the model did not extract the crew 2 launch event. 
However, this might be explained by the fact that the digital trends news outlet saw a sharp decline in articles output during this period as can be seen in figure \ref{articletime}.

Now, we will look at the document tree for one document, see figure \ref{doctree}. 
This was created by choosing only the topics that were assigned to at least 5\% of words in the chosen document. 
The document in question is titled "Astronauts are using VR to train for the Boeing Starliner capsule". 
The three main extracted topics are virtual reality applications, space and R\&D. 
Two children of the topic of space were also assigned to this document: test and astronauts. 
From the title, it can be seen that the tree captures the main themes of this document.

\section{Conclusion}

We have proposed a new model for topic modelling capable of modelling hierarchy and time jointly. 
Through examples, we have demonstrated how combining hierarchy and temporality provides us with a more fine grained understanding of a corpus through detailed sub-topics which can represent specific events.
Moreover, we developed a novel implementation of Gibbs sampling for hierarchical topic models. 
This implementation provides a fast alternative to SVI that makes Gibbs sampling a viable solution for training such complex models. 
Moreover, we have shown how extracting entities can help interpret and understand topics at a deeper level. 

\section{Limitations}
Our model is subject to a few limitations that lays the foundations for future work.
We have inherited the general limitations existing in all topic models techniques.
\begin{itemize}
\item As topic models require tokenization, non-tokenizable languages like Chinese are not compatible
\item Since we cannot be aware of all the content of the training corpus, it is difficult to determine if some topics were missed during extraction; we can only evaluate the topics that are extracted.
\item Hyper parameters defining priors on probability distributions may depend on the specifics of the dataset such as the number of articles, their average length or how varied/narrow a corpus is (affecting the number of topics to expect)
\item Topics must be interpreted by humans which is not always a simple task even with additional information such as top entities or top documents
\end{itemize}
We also have  limitations that are specific to our method
\begin{itemize}
\item Convergence must be observed to confirm the end of the training phase. However, as hierarchical topic model can extract hundreds of topics, we cannot ensure the convergence of each topic manually and only asses the first level topics.
\item The deeper a topic is, the more esoteric it becomes. Hence, it can be difficult to interpret such topics as it require specific domain knowledge. 
\item Since we cannot be aware of all the content of the training corpus, it is difficult to determine if some events (topics localized in time) were missed during extraction. we can only evaluate the events that are extracted.
\end{itemize}



\bibliographystyle{META/acl_natbib}
\bibliography{META/bib}

\appendix

\section{Tables and figures}

\begin{figure}[ht]
    \centering
    \includegraphics[angle=90,width=.45\textwidth]{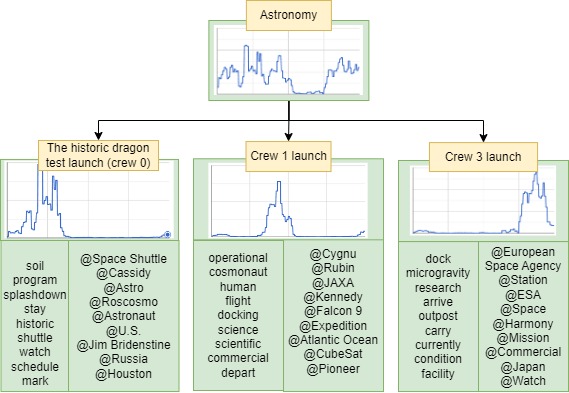}
    \caption{Examples of depth 3 topics that are well localized in time}
    \label{tempo}
\end{figure}

\begin{figure}[ht]
    \centering
    \includegraphics[width=.5\textwidth]{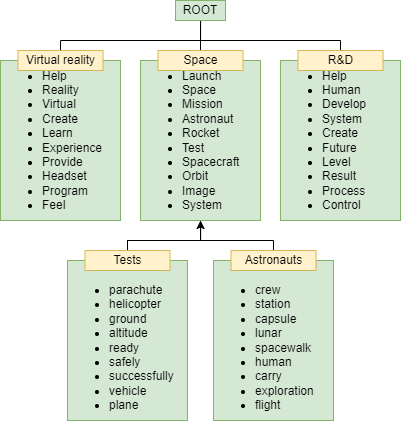}
    \caption{Example of document topic tree for the document : "Astronauts are using VR to train for the Boeing Starliner capsule" .}
    \label{doctree}
\end{figure}

\begin{figure}[ht]
    \centering
    \includegraphics[width=.5\textwidth]{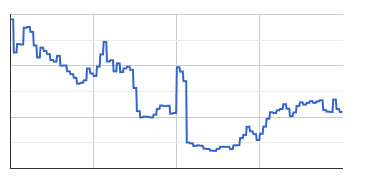}
    \caption{Number of articles published by Digital Trends over the years 2020 and 2021. We can see a sharp decline at the beginning of the year 2021 (middle of the graph)}
    \label{articletime}
\end{figure}

\begin{figure*}[ht]
    \centering
    \includegraphics[width=0.9\textwidth]{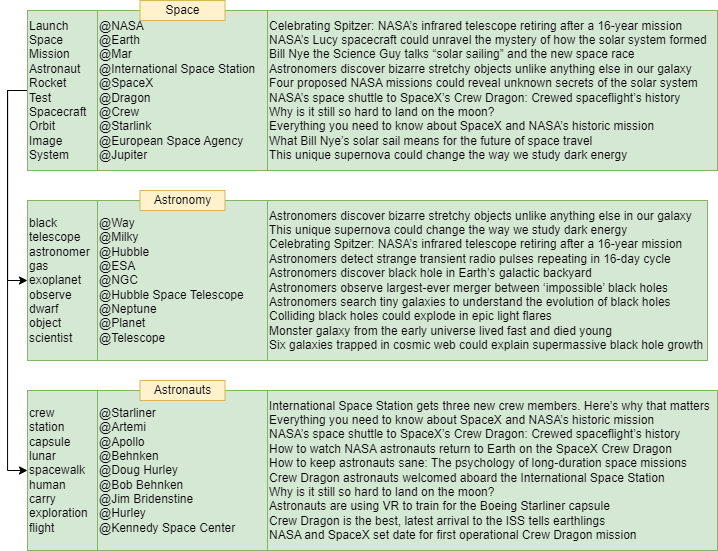}
    \caption{The space topic and two of its sub-topics : astronauts and astronomy. Each topics is shown with top words (left), top entities (center) and top documents (right)}
    \label{space}
\end{figure*}

\begin{table*}[ht]
\centering
\begin{tabular}{lll}
\hline
\textbf{Name}                        & \textbf{Value}                 & \textbf{Category}                                              \\ \hline
$\alpha$                    & 0.00005               & \multirow{3}{*}{IDTs parameters}      \\ 
$\beta$                     & 0.0002                &                                                   \\ 
$\theta$                    & 0.25                   &                                                   \\ \hline
Critical Mass (CM)          & 0.0005                & \multirow{3}{*}{IDTs growth control}                   \\ 
Splitting Mass (SM)         & 0.005                 &                                                   \\ 
Time To Live (TTL)          & 2                     &                                                   \\ \hline
$\phi$                      & 0.1                   & \multirow{2}{*}{Traditional LDA topic parameters} \\ 
$\epsilon$                    & 1                     &                                                   \\ \hline
Iterations                  & 4500                 & \multirow{2}{*}{Training parameters}              \\ 
Batch size                  & 500                   &                                                   \\ \hline
\end{tabular}
\caption{Parameters used for our model}
\label{tab:param}
\end{table*}

\begin{table*}[ht]
\centering
\resizebox{\textwidth}{!}{%
\begin{tabular}{llll}
\hline
\textbf{Model name} & \textbf{Type}         & \textbf{Reference}           & \textbf{Corpora}                                       \\ \hline
LDA        & Classic      & \citep{LDA}         & News articles                                 \\ 
HDP        & Classic      & \citep{HDP}         & Scientific papers                             \\ 
nCRP       & Hierarchical & \citep{nCRP}        & Abstracts                                     \\ 
PAMmix     & Hierarchical & \citep{PAMmix}      & Abstracts                                     \\ 
nHDP       & Hierarchical & \citep{nHDP}        & News articles and Wikipedia                   \\ 
LSHTM      & Hierarchical & \citep{LSHTM}       & News articles and Wikipedia                   \\ 
DTM        & Temporal     & \citep{DTM}         & Scientific papers                             \\ 
ToT        & Temporal     & \citep{ToT}         & Scientific papers, mails and historical texts \\ 
MTT        & Temporal     & \citep{MST}         & News articles                                 \\ 
DCTM       & Temporal     & \citep{song2008non} & Scientific papers                             \\ \hline
\end{tabular}%
}
\caption{Related methods corpora, evaluation methods used and type of topic model.}
\label{tab:related}
\end{table*}
\begin{table*}[ht]
\centering
\begin{tabular}{ll}
\hline
\textbf{Variable}                  & \textbf{Description}                                               \\ \hline

$A^*(z_{0,j})$ & \# words assigned to topic $z_{0,j}$  \\ 
P                         & Weight of the currently selected node $z_{0,j}$.              \\ 
C                         & Weight of all of the children of the selected node $z_{0,j}$. \\ 
N                         & Weight of a potentially new child for $z_{0,j}$              \\ 
$\theta_1$ and $\theta_2$ & Prior for the Bernoulli distribution                      \\ \hline
\end{tabular}
\caption{Descriptions of variables for equations \ref{bern1} to \ref{bernn}}
\label{tab:bernoulli}
\end{table*}

\begin{table*}[ht]
\centering
\resizebox{\textwidth}{!}{%
\begin{tabular}{ll}
\hline
\textbf{Variable} & \textbf{Description}                                 \\ \hline
$n$      & \# words in the corpus               \\ 
$n_d$    & \# words in the corpus that are part of document $d$           \\ 
$n_w$    & \# words in the corpus that are instantiations of the word $w$ \\ 
V                         & Vocabulary length                                         \\ 
$A(k |w)$     & \# words $w$ assigned to topic $(z_{0,j-1},k)$ or its descendants (corpus tree information)               \\ 
$A(k |d)$     & \# words in document $d$ assigned to topic $(z_{0,j-1},k)$ or its descendants (document tree information) \\
$A(k)$     & \# words assigned to topic $(z_{0,j-1},k)$ or its descendants               \\  
$\beta_{k}$    & Probability density function of the beta distribution with parameter $\rho^1_k$ and $\rho^2_k$ associated with topic $(z_{0,j-1},k)$                      \\ 
$\epsilon, \phi, \beta, \alpha$ & Priors for the Dirichlet distributions and processes (more details are provided in the parameter section)           \\ \hline
\end{tabular}%
}
\caption{Descriptions of variables for equations \ref{eq:sampling} to \ref{eq:samplingN}}
\label{tab:samplingz}
\end{table*}
\end{document}